\documentclass{ws-procs961x669}            
\usepackage{siunitx}
\begin{document}
\title{Solstices and Equinoxes in 1703 at the meridian line of St. Maria degli Angeli in Rome, and the stellar aberration of Sirius}

\author{Costantino Sigismondi$^1$}

\address{ICRA, Sapienza University of Rome,\\
Rome, Italy\\
$^1$E-mail: sigismondi@icra.it\\
 \url{http://www.icra.it/solar}}
 
 \author{Silvia Pietroni$^2$}

\address{ \quad Dipartimento di Fisica “E.R. Caianiello”, Università di Salerno\\ Via Giovanni Paolo II 132,
I-84084 Fisciano, Italy\\
 \quad  Istituto Nazionale di Fisica Nucleare, Sezione di Napoli\\ Via Cintia, 80126, Napoli, Italy\\
$^2$E-mail: spietroni@unisa.it}

\begin{abstract}

The 1703 was the first year of full operation of the meridian line in the Basilica of St. Maria degli Angeli in Rome. The instants of solstices and equinoxes, the {\sl{Anni Cardines}}, obtained by comparing transit timings of Sun and Sirius, also in daytime, are affected either by the East deviation of the meridian line of about 5' Eastward (geometrical effect), either by the stellar aberration of Sirius (relativistic effect). 
Similarly the seasonal shifts of Sirius' declination observed by Bianchini are here firstly recognized as depending on the stellar aberration in ecliptic latitude.
The eccentricity of the Earth's orbit and the orientation of its axis in the space can be measured, since Ptolemy, by knowing the exact timing of the solstices and the equinoxes.
The dates of 1703 equinoxes and solstices have been published by Francesco Bianchini in local roman mean time, referred to the local (roman) mean noon, i.e. after the solar meridian mean transit time. 
By using the observations of the three lunar eclipses of 1703 we found the equation of time for that year, and the UTC corresponding timings of the beginning of the seasons. This operation lead to find the contribution of the Sirius aberration to the instant calculated by Bianchini with respect to the current celestial mechanical models of IMCCE. 

\end{abstract}


\bodymatter

\section{Stellar aberration measures in 1701 before Bradley's discovery}
The stellar aberration was discovered by James Bradley in 1727, and the same astronomer discovered the nutation of the Earth's axis in 1737. At the meridian line of St. Maria degli Angeli, the giant Clementine Gnomon, built by astronomer Francesco Bianchini (1662-1729) and funded by the Cardinal Gianfrancesco Albani (1649-1721), elected pope on 23 November 1700 with the name of Clement XI. 

The effects of stellar aberration have been detected on the Polaris (2006), and on Sirius (2021, present work).

The aberration of the Polaris influenced the measure of the latitude of the pinhole, the gnomon of the meridian line, made by Bianchini on January 1-8, 1701.

The aberration on Sirius affected the instant of the equinoxes and the solstices, calculated with the difference between solar and stellar meridian transit, the latter observed also in full daylight. To obtain such evidence of the first special relativistic effect observable, a complete calibration of the meridian line has been carried since 2018. With the IGEA observational campaign, Informatized Geometric Ephemerides for Astrometry:
all the reference points present of the 45 meters long meridian line have been calibrated, by comparing solar observations and ephemerides. 

At the Marcel Grossmann Meeting XI (Berlin, 2006) the evidences of Polaris' aberration were discovered in the latitude, from the measures by Bianchini at the Gnomon, and now at the Marcel Grossmann XVI edition we can afford the evidences on Sirius' stellar aberration with the increased precision on the calibration of this giant instrument, the second {\sl{Heliometer}}, or solar telescope, of its times, after the one of Cassini in St. Petronio, Bologna.

\section{Managing dates and timing of 1703, in local mean time}

The equation of time shifts the instants of the meridian transit day by day with respect  to an absolute time reference such as UTC. The eclipses' ephemerides are based upon the terrestrial dynamical time TDT, while the terrestrial rotational time UT1 includes the slowing down effect of the Earth's rotation in act since 1700. To translate 1703 local time to modern UTC to the nearest second we exploited the observations of the three lunar eclipses occurred in 1703, all made by Bianchini. The equation of time of that year come out from the comparison between the observed instants of eclipse's beginning and end -in local mean time- and the corresponding instants in TDT computed by the NASA ephemerides.

\section{The dates and timing for 1703 lunar eclipses from NASA ephemerides}

The three lunar eclipses of 1703 have been observed by Francesco Bianchini and published in the same book (Bianchini, 1703) where the great meridian line, the {\sl{Clementine Gnomon}} was presented.
The instants of beginning and end of lunar eclipses are independent of the position of the observers on the Earth, and they can be calculated with a suitable celestial mechanical model of the lunar orbit (provided by NASA).
They occurred very close to the solstices, making the procedure of their absolute timing (expressed in TDT and then in UTC) particularly accurate.

\begin{figure}[htbp!]
\centering
\includegraphics[width=8 cm]{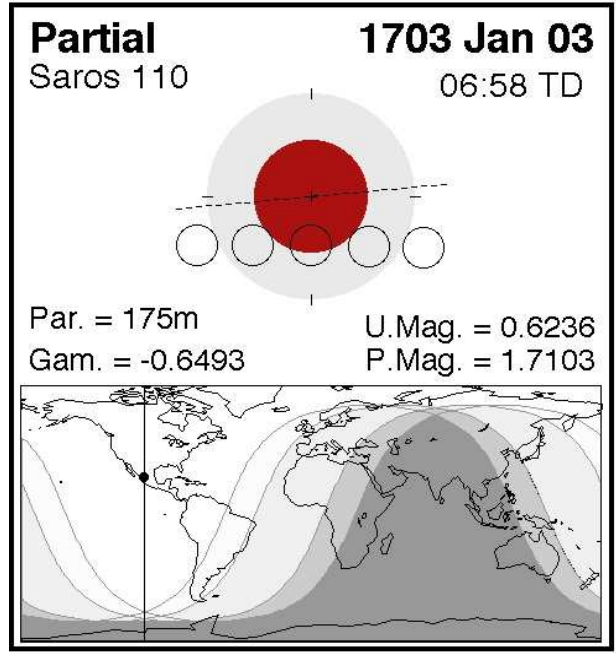}
\caption{The lunar eclipse of January $3^{rd}$ 1703}
\label{aba:fig1}
\end{figure}

\begin{figure}[h]
\centering
\includegraphics[width=8 cm]{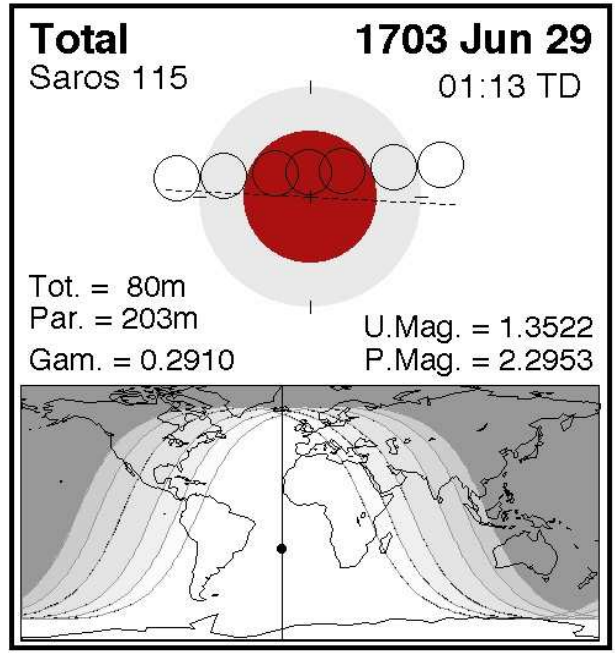}
\caption{The lunar eclipse of June $29^{th}$ 1703}
\label{aba:fig2}
\end{figure}

\begin{figure}[htbp!]
\centering
\includegraphics[width=8 cm]{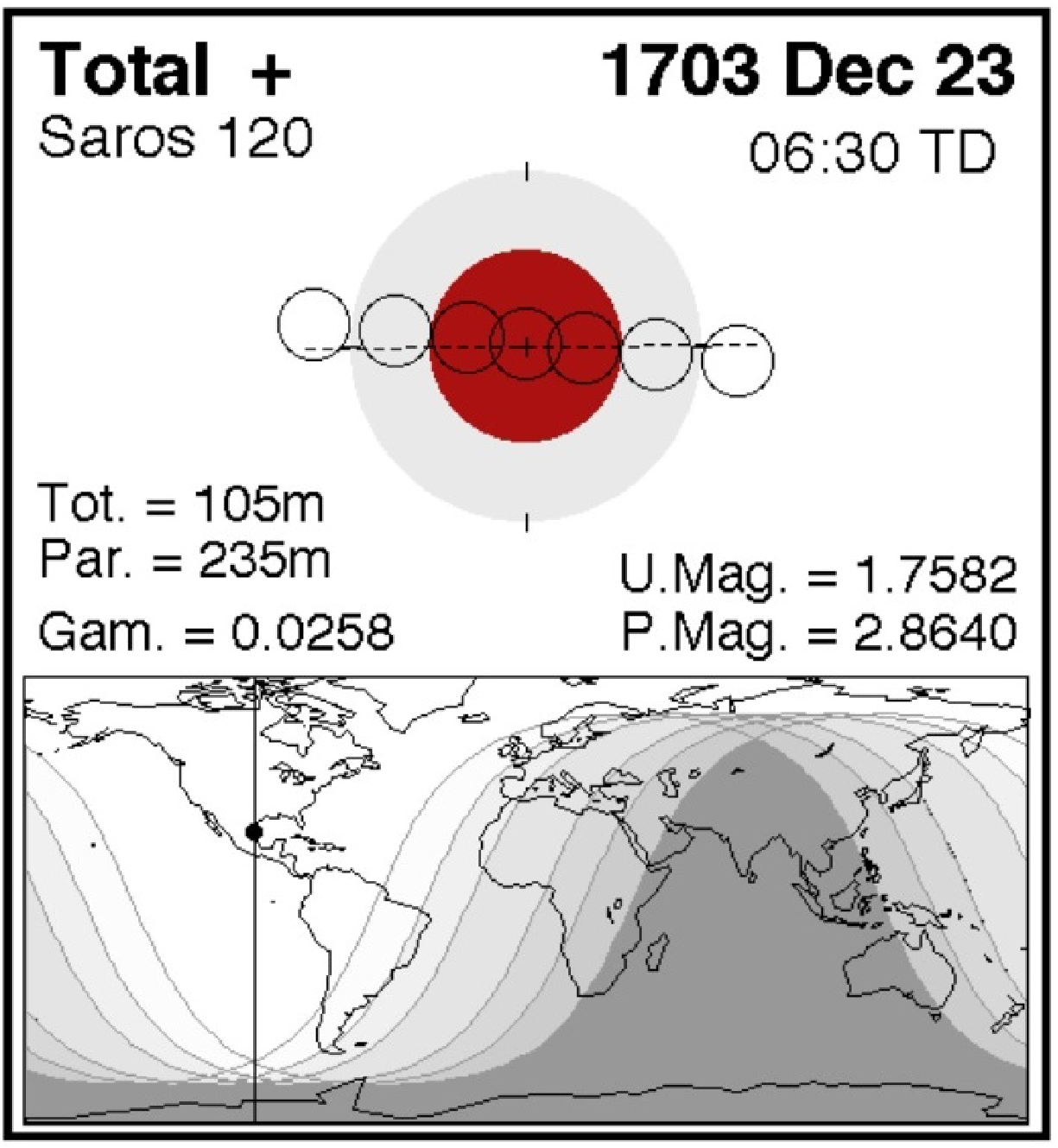}
\caption{The lunar eclipse of December $23^{rd}$ 1703}
\label{aba:fig3}
\end{figure}

The maximum phase of the eclipse is indicated in TD, terrestrial dynamical time to the nearest minute in the figures, and to second accuracy in the website. The maximum of the eclipse is recovered in local time after doing the average between starting time and ending time of the eclipse, to the nearest second.
In the details of the  {\sl{Canon of five millennia of lunar eclipses}}, also the starting time and the ending time of the eclipses are provided.
The eclipses have been observed by Bianchini at the telescope, improving the accuracy of the measurements.

\section{Solstices and equinoxes and the Equation of time in the ephemerides}

The dynamical model for the Earth's orbit includes the perturbations of all planets, and they are implemented in the most advanced ephemerides.
IMCCE provides an online service back to 4000 b. C. for the dates of the astronomical beginning of the seasons.
The comparison with the corresponding instant of 1703 as calculated by Bianchini required a conversion into UTC of the local timings. 
It is available an online e-service, {\sl{Planetcalc}}, with the equations of time, from 500 b. C.

\begin{figure}[htbp!]
\centering
\includegraphics[width=11 cm]{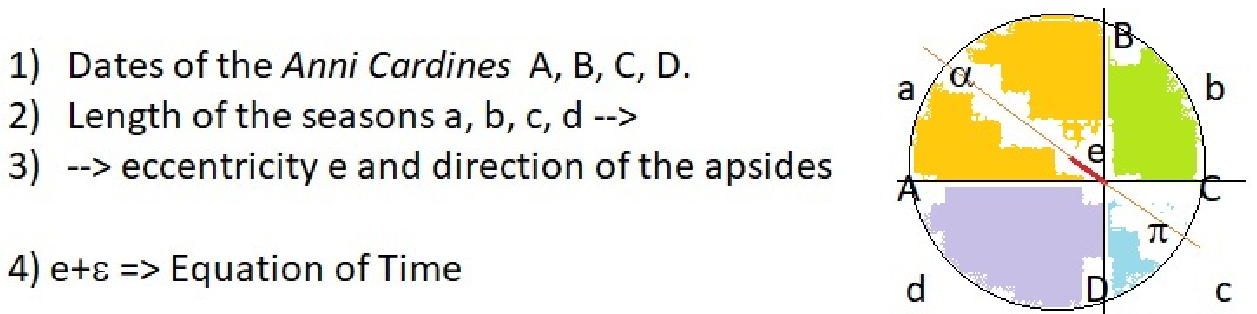}
\caption{The connection between the beginning of seasons' times and the eccentricity, and the equation of time. Aphelion $\alpha$ and perihelion $\pi$, the apsides, are also sketched.}
\label{aba:fig4}
\end{figure}

\begin{figure}[htbp!]
\centering
\includegraphics[width=11 cm]{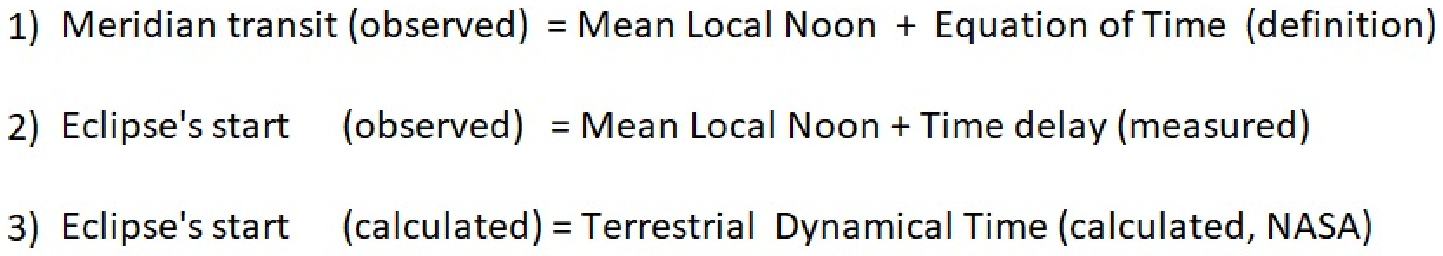}
\caption{The relations existing between Mean Noon, Equation of time, solar transit time and terrestrial dynamical time TD.}
\label{aba:fig5}
\end{figure}

We need a self-consistent derivation of the UTC timing of the equinoxes and the solstices in 1703, being the Equation of Time itself depending on the Earth's orbit eccentricity, which is derived from the the observed beginning of the seasons, namely the equinoxes and solstices.

\begin{figure}[htbp!]
\centering
\includegraphics[width=11 cm]{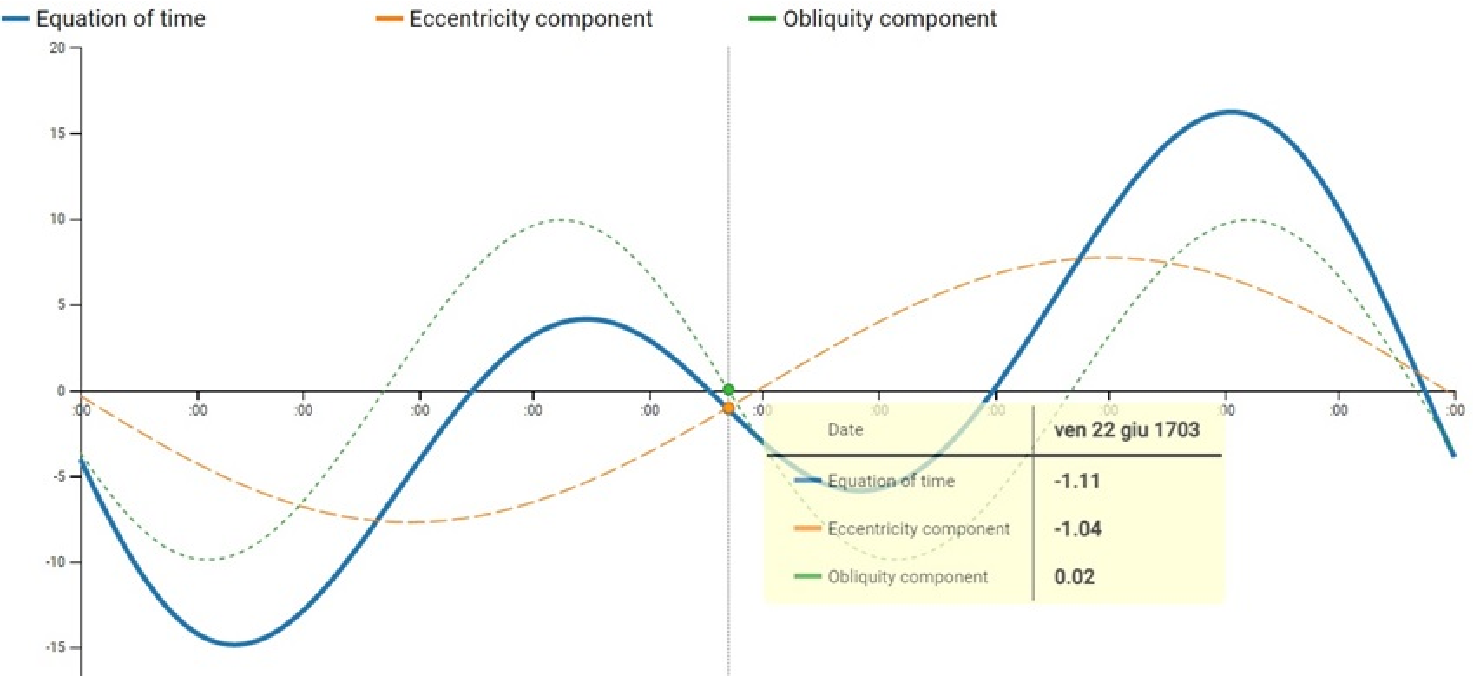}
\caption{The equation of time for 1703, as computed with Planetcalc.}
\label{aba:fig6}
\end{figure}

The obliquity component goes to zero four times per year at the equinoxes and the solstices, while the eccentricity component vanishes twice per year at the apsides (aphelion and perihelion). This is well shown in {\sl{Planetcalc}} website.

\section{Synchronization of the Equation of time with lunar eclipses in 1703}

The dates of the apsides (as shown in Sigismondi and Pietroni, 2019) and the obliquity of the year are used to compute the equation of time for that year, but an independent synchronization with UTC  is needed, to verify the IMCCE ephemerides, and the lunar eclipses were used exactly to assure that synchronization.
On June $22^{nd}$ 1703 the equation of time of Planetcalc is $-1.11$ minutes, and in $28^{th}$ June, last solar transit before the lunar eclipse, it is $-2.38$ minutes.
It means that this local noon should occur 2 m 22 s before a mean solar noon fixed (e.g. CET Central European Time ).
For Rome the mean noon is at 12:10 CET, then 12:12:22 the transit time for $28^{th}$ June, 1703.
The eclipse was computed as maximum at 1:13 TD on $29^{th}$ June, so 2:13 CET.
Bianchini's beginning of totality is at 13:18:15 after solar meridian transit of $28^{th}$ June, and the end (emersion) is 14:38:15.
The middle of totality (maximum phase) is at 13:28:15, obtained by averaging.
The solar transit in the same publication occurred at 12:07:36
so that the eclipse's maximum was at 02:05:51.
The terrestrial dynamical time for the eclipse has been computed (Espenak and Meeus, 2009) on 02:12:47 (1:13 TD in figure 1). The TD, is ahead with respect to Bianchini's roman mean time of 6 min 56 s.

\section{Including $\Delta UT1$ in the final timing}

The procedure shown in detail for June $29^{th}$ has been applied to the other eclipses of 1703, that Bianchini observed also with Filippo Maraldi (January $3^{rd}$) and alone (December $23^{rd}$), recovering the Roman average time adopted by Bianchini for his observations, with respect to the Terrestrial Dynamical Time.
The adopted value for $\Delta$UT1, of 7-8 seconds for the 1703, and due to the de-rotation effect of the Earth upon continental uplift after the ice age, is added to TD to obtain the coordinate universal time UTC for these observations, and for the dates and timing of solstices and equinoxes in 1703.
These instants of seasons' beginning dates have been used for studying the aberration effect on the solstices and equinoxes in 1703 at the meridian line of St. Maria degli Angeli in Rome, demonstrating that Sirius' aberration and the deviation from Celestial North of the meridian line have caused the departure from IMCCE calculated times for 1703 seasons' starts (Sigismondi, 2021).

\section{The dates and timing for 1703 \sl{anni cardines}}

Bianchini used the term {\sl{Anni Cardines}} for the solstices and the equinoxes, because they are related to:
\begin{itemize}
    \item Inclination of the Earth's axis on the ecliptic plane (the plane of the orbit), called {\sl{obliquity}};
      \item Position of the apsides in the Earth's orbit, which determines the {\sl{eccentricity}}.
\end{itemize}

They were published in the book {\sl{De Nummo et Gnomone Clementino}} of 1703, as well as on a marble epigraph in the presbytherium.

\begin{figure}[htbp!]
\centering
\includegraphics[width=12 cm]{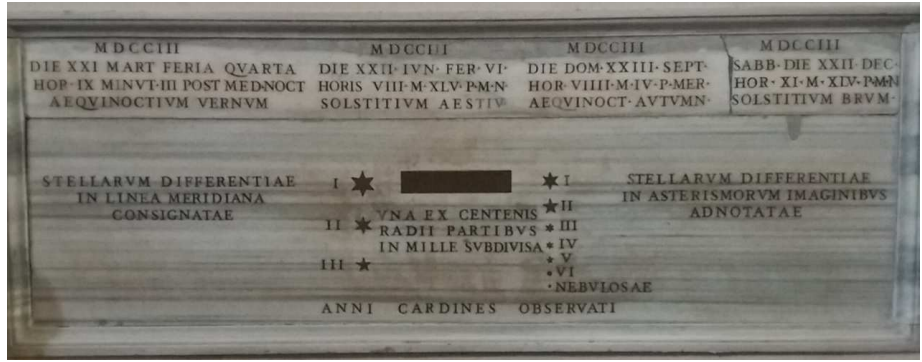}
\caption{The epigraph of 1703 {\sl{Anni Cardines}} in the presbytherium of St. Maria degli Angeli in Rome.}
\label{aba:fig7}
\end{figure}

These instants are expressed in mean local time, after/before the mean meridian transit or midnight (anti-)transit.
So to recover their instants as expressed in UTC it was necessary to recover three lunar eclipses observed in the same year and published by Bianchini in the same book (Bianchini, 1703): January 3, June 29 and December 23 (see below).
After this operation of synchronization, we made the comparison with the IMCCE ephemerides for 1703.

\begin{figure}[htbp!]
\centering
\includegraphics[width=12 cm]{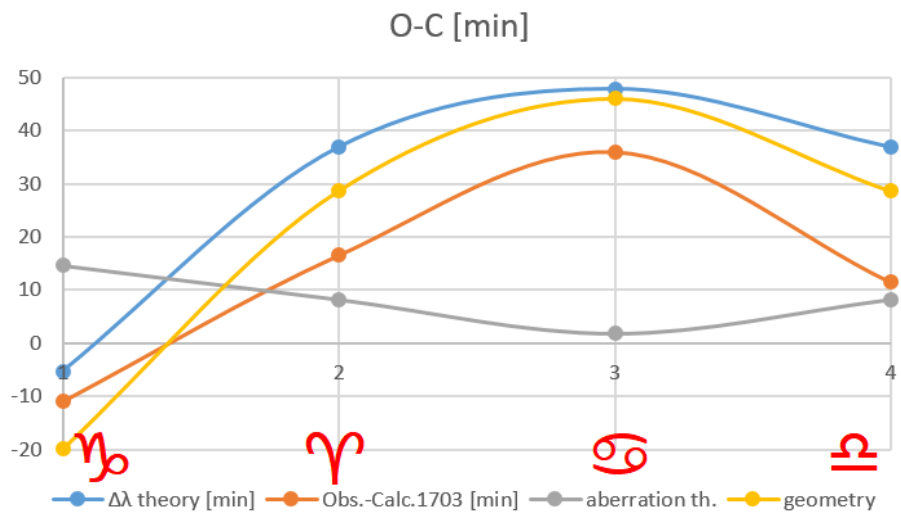}
\caption{The epigraph's timings translated into UTC and compared with IMCCE ephemerides.}
\label{aba:fig8}
\end{figure}

\newpage

\section{Sirius' aberration and the East deviation of the line}

Before the present analysis it was not clear that Bianchini used only Sirius for timing the solar meridian transits, also in daytime.
The reason can be practical: Sirius is the brightest star in the sky and it can be seen in full daylight with a telescope through the open window 60 cm $\times$ 40 cm located over the pinhole.
I made personally observations of meridian transits of Antares, Ras Alhague and Venus in daytime, outdoor at the meridian line of the Vatican obelisk, and the brightness of the star is crucial.
Nevertheless on the meridian line of St. Maria degli Angeli there are reported 22 stars, with their ecliptic coordinates of 1701. This lead us to intend that all the 22 stars were used during the measurements (Bedinsky and Nastasi, 2007), while only Sirius was certainly used, after the examination of the documents of Bianchini.

The {\sl{Anni Cardines}} are with ecliptic longitude $\ang{0}$, $\ang{90}$, $\ang{180}$, $\ang{270}$. 

Sirius in 
1701 had ecliptic longitude $\ang{97} 57' 53"$ (fig. 3), and it was supposed to change only for the precession, until James Bradley discovered the aberration in 1727.

\section{Seasonal aberration in ecliptic latitude}

Francesco Bianchini (1703) considered also the meridian shift of Sirius, as of seasonal and of meteorological cause.  Here I have demonstrated that it was due to the stellar aberration in ecliptic latitude (being Sirius not on the ecliptic plane but nearly at $\ang{-40}$).

\begin{figure}[htbp!]
\centering
\includegraphics[width=10 cm]{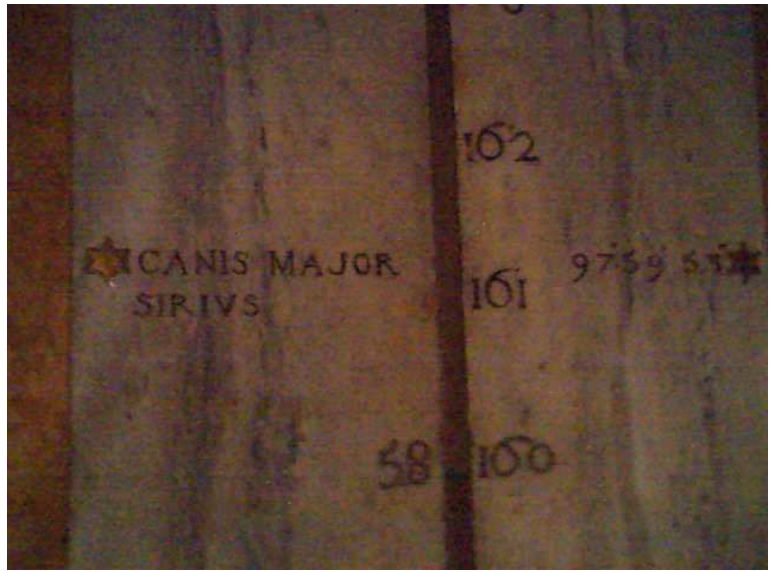}
\caption{The coordinates of Sirius in 1701 from the Stellar Atlas of Philippe de la Hire, as reported on the meridian line of St. Maria degli Angeli.}
\label{aba:fig9}
\end{figure}

\begin{figure}[htbp!]
\centering
\includegraphics[width=13 cm]{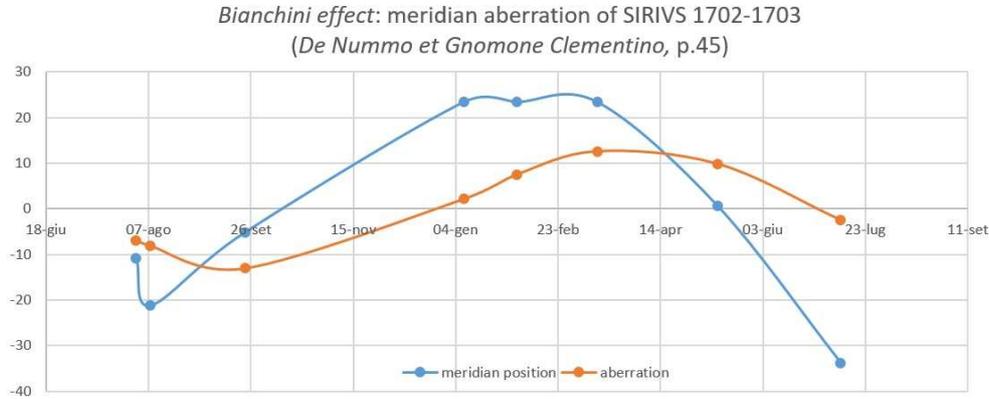}
\caption{The meridian position of Sirius along the year. Bianchini thought it was a meteorological effect, while it is due to aberration in ecliptic latitude.}
\label{aba:fig10}
\end{figure}

\section{Discussion}

The measurements reported by Bianchini on the epigraph are plotted in orange, and they are in very good agreement with the blue line which includes the effects of the line's deviation (geometry in yellow) and of the stellar aberration in grey.
The geometrical effect is much larger, nevertheless the aberration's contribution matches exactly as needed, especially in the cases of opposition (Sun in Capricorn) and conjunction (Sun in Cancer) with the star.
The accuracy of the method applied at the Clementine meridian line is enough to recover the {\sl{Anni Cardines}} within 10 minutes, as Bianchini declared in his book.
The departure from this value, up to 30 minutes, is due to a systematic error (the deviation of the whole line toward East of 5.2', or 70 mm at its Northern extreme, 45 meters from the pinhole's vertical) combined with the special relativistic effect of the stellar aberration of Sirius, used in all measurements.

\section{Conclusions}

An important contribution of this paper is the verification that Sirius was the star always used in the meridian measurements, to compute the ecliptic coordinates of the Sun with respect to a fixed star.
Finally the effect of meridian shift of Sirius along the year, was clearly recognized by Bianchini as seasonal, and it is, not because of seasonal humidity variations, but of special relativistic origin: due to the stellar aberration in ecliptic latitude, which is a rather big effect being Sirius of  ecliptic latitude $b=\ang{-40}$.

The operation of synchronizing the mean solar time of Rome in 1703 with the Terrestrial Dynamical Time and then with the Universal Time Coordinated, has made possible recovering the Equation of Time of 1703 and the comparison between modern ephemerides and these data, to distinguish clearly the stellar aberration's contribution of Sirius, from the one due to the Eastward deviation of the line.

\end{document}